\documentclass{ws-ijmpd}
\begin{document}

\title{Vacuum pressures and energy in a strong magnetic field}

\author{H. P\'erez Rojas and E. Rodr\'iguez Querts}

\address{Instituto de Cibern\'etica, Matem\'atica  y
F\'{\i}sica,\\ calle E esq. a 15 No. 309.Vedado, C. Habana,Cuba
\\ hugo@icmf.inf.cu, elizabeth@icmf.inf.cu}


\maketitle

\begin{abstract}We study vacuum in a strong magnetic field. It
shows a nonlinear response, as a ferromagnetic medium. Anisotropic
pressures arise, and a negative pressure is exerted in the
direction perpendicular to the field. The analogy of this effect
with the Casimir effect is analyzed. The vacuum transverse
pressure is found to be of the same order of the statistical
pressure  for $B\sim10^{15}G$ and $N\sim10^{33}electrons/cm^{3}$.
Vacuum interaction with the field is studied also for
$B\sim10^{16}G$ and larger, including the electron anomalous
magnetic moment. We estimate quark contribution to vacuum
behavior. \end{abstract}
\section{Introduction}

\indent

There is an analogy among certain boundary conditions and the
effect of external fields. Some boundary conditions lead to new
physical effects as it is for instance the anisotropic box created
by two parallel metallic plates of length $a$ separated by a
distance $d$, placed in vacuum. For $d<<a$ an attractive force
appears in between the plates, as produced by the zero point
vacuum energy of electromagnetic modes, leading to the well-known
Casimir effect \cite{Casimir},\cite{Milonni}.

The Casimir effect shows that if one breaks the symmetry in a
region of space, the energy $E_V$ of the vacuum modes results
distributed anisotropically. We may consider the vector ${\bf d}$
perpendicular to the plates as characterizing the symmetry
breakdown. Only vacuum modes of momentum $p_{nd}=\frac{2\pi n
\hbar}{d},n=1,2,...$ in the direction of symmetry breakdown  are
allowed inside the cavity (then
$E_V=\sqrt{p_1^2+p_2^2+p_{nd}^2}$). A negative pressure dependent
on $d$ arises inside the plates, and perpendicular to them.

On the other hand, it is known that in an external magnetic field
$B$, the momentum
 of an electron (positron) in the direction perpendicular to
 the field is quantized $p_{\perp}=\sqrt{2eBn}$. It is interesting to
inquire then if an effect similar to the Casimir one is produced
by the zero point electron-positron energy of vacuum in an
external constant magnetic field.

\section{Vacuum properties in a constant magnetic field}

\indent

\begin{figure}[t]
\epsfxsize=25pc 
\epsfbox{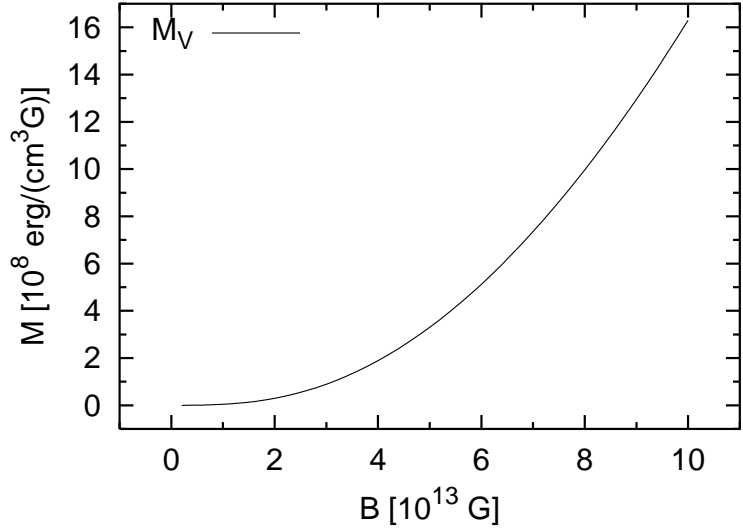} 
\caption{Vacuum magnetization ${\cal M}_{V}$ for magnetic fields
$B\sim10^{13}G$.
 \label{fig:magnetization}}
\end{figure}

The electron-positron zero point vacuum energy in an external
electromagnetic field was obtained by  Heisenberg and Euler
\cite{Euler}. In the specific case of an external constant
magnetic field, this expression results as the pure vacuum term
contribution  when calculating the tadpole term of the
thermodynamic potential $\Omega$ in a medium \cite{h plasma}. In
this calculation one starts from the energy eigenvalues of the
Dirac equation for an electron (positron) in a constant magnetic
field $B$,
\begin{equation}\label{eigenv}
  \varepsilon_{n}=\sqrt{p_{3}^{2}+m^{2}+2eBn},
\end{equation}
where $ n=0,1,2,...$, are the Landau quantum numbers, $p_{3}$ is
the momentum component along the magnetic field (we consider $B$
parallel to the third axes) and $m$  is the electron mass. The
general expression for $\Omega$ contains two terms
\begin{equation}
  \Omega=\Omega_{ST}+\Omega_{V}.
\end{equation}
The first one is the quantum statistical contribution which
vanishes in the limit of zero temperature and zero density. The
second term accounts for the zero-point energy: it contains the
contribution coming from the virtual electron-positron pairs
created and annihilated spontaneously in vacuum and interacting
with the field $B$. In the one loop approximation, where no
radiative corrections are considered, it has the expression
\begin{equation}\label{pot1}
\Omega_{V}=-\frac{e
B}{4\pi^{2}}\sum_{n=0}^{\infty}\alpha_{n}\int_{-\infty}^{\infty}dp_{3}\varepsilon_{n},
\end{equation}
with $\alpha_{n}=2-\delta_{n0}$.  As can be observed, (\ref{pot1})
is a divergent quantity. After regularization it leads to the
 Euler-Heisenberg
expression
\begin{equation} \Omega_{V} =
\frac{e^2 B^2}{8\pi^2}\int_0^{\infty}e^{-m^2 x/eB}\left[\frac{coth
x}{x} -\frac{1}{x^2}-\frac{1}{3}\right]\frac{d x}{x}. \label{EH}
\end{equation}
We observe that the vacuum term regularization demands the
addition of a negative infinite term proportional to $B^2$ which
absorbs the classical energy term $B^2/8\pi$.
 The vacuum thermodynamic potential is actually negative. We
  interpret it according to the general energy-momentum tensor
expression in a constant magnetic field (see i.e. \cite{h3}) for
the case of zero temperature and zero chemical potential,
\begin{equation}\label{KSP}
{\cal T}_{0 \mu,\nu}=  4F_{\mu \rho} F_{\nu \rho}\partial
\Omega_{V}/\partial F^2 - \delta_{\mu \nu}\Omega_{V}, \label{KSP}
\end{equation}
leading to a positive pressure $P_{V 3} = -\Omega$ along the
magnetic field $B$, and to a negative pressure $P_{V \perp}=
-\Omega-B{\cal M}$ in the direction perpendicular to the field,
where ${\cal M}_{V}=-\frac{\partial\Omega_{V}}{\partial B}$ is the
magnetization, which is obtained from (\ref{EH}) as
\begin{equation}\label{magnetization}
{\cal M}_{V}=-\frac{2\Omega_{V}}{B}-
\frac{em^2}{8\pi^2}\int_0^{\infty}e^{-m^2 x/eB}\left[\frac{coth
x}{x} -\frac{1}{x^2}-\frac{1}{3}\right]d x
\end{equation}
\noindent It is easy to see that the magnetization
(\ref{magnetization}) is a positive quantity. Moreover, it has a
non- linear dependence on the field $B$, as is shown in Fig.
\ref{fig:magnetization} . In this sense, vacuum has ferromagnetic
properties, although in our present one-loop approximation we do
not consider the spin-spin interaction between virtual particles.

 Concerning the transverse
pressure $P_{V \perp}=-\Omega -{\cal M}B$, we get
\begin{equation}\label{pressure}
P_{V \perp}=\Omega_{V} + \frac{m^2 e
B}{8\pi^2}\int_0^{\infty}e^{-m^2 x/eB}\left[\frac{coth x}{x}
-\frac{1}{x^2}-\frac{1}{3}\right]d x.
\end{equation}
$P_{V \perp}$ is negative and it may lead to some effects for
small as well for high fields. It must be stressed that the term
$B{\cal M}$ subtracted to $-\Omega$ in $P_{V \perp}$ is the
quantum statistical analogue of the pressure due to the Lorentz
force for particles (in the present case virtual) bearing a
magnetic moment, which leads to ${\cal M}>0$ \cite{Aurora}

\section{The low energy limit $eB<<m^{2}$}

\indent

 Fields currently achieved in laboratories are very small
if compared with the critical field $B_{c}=m^{2}/e\sim10^{13}$.
When this limit condition holds $B<< m^2/e$, one can write,
\begin{equation}\label{magnetization2}
{\Omega_{V}}\approx-\frac{(eB)^{4}}{360\pi^{2}m^{4}},
\end{equation}
 and in consequence
\begin{equation}
 P_{V \perp} \approx
-\frac{(eB)^{4}}{120\pi^{2}m^{4}}.
\end{equation}
In usual units it reads
\begin{equation}\label{pv}
  P_{V \perp}\approx
-\frac{\pi^{2} \hbar c}{120b^{4}},
\end{equation}
 where the characteristic parameter $b(B)$ is
\begin{equation}
  b(B)= \frac{2 \pi \lambda_{L}^{2}}{\lambda_{C}}.
\end{equation}
 Here, $\lambda_{L}$ is the magnetic wavelength
\begin{equation}
  \lambda_{L}=\sqrt{\frac{hc}{2eB}}
\end{equation}
  and $\lambda_{C}$ is the Compton wavelength $\lambda_{C}=\frac{h}{mc}$. It is easy to see
  that $\lambda_{L}^{2}$ coincides with the area corresponding to one magnetic
  flux quantum $\Phi_{o}=\frac{hc}{2e}$. Pressure then is a
  function of
  the field dependent parameter $b(B)$, which is determined by the
  ratio between the Compton wavelength and the "one fluxon area".

  The expression for the transverse pressure (\ref{pv}) looks similar to
  the expression for the negative pressure due to the Casimir effect between parallel
metallic plates \cite{Casimir}
\begin{equation}\label{Cassimir}
 P_C=-\frac{\pi^2 \hbar c}{240 d^4}.
\end{equation}
\begin{figure}[t]
\epsfxsize=25pc 
\epsfbox{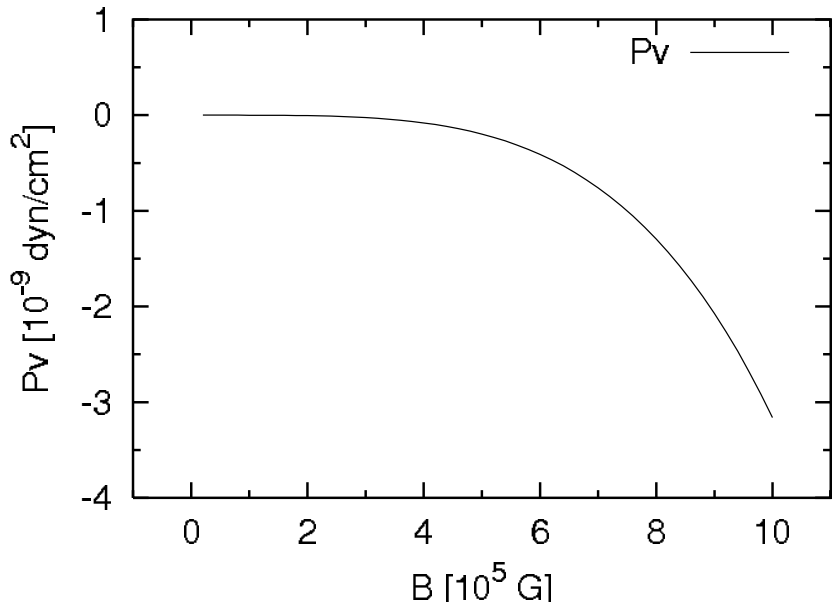} 
\caption{Vacuum transverse pressure for small fields
$B\sim10^{5}G$. \label{fig:casimir}}
\end{figure}
$P_{V \perp}$ is then a Casimir- like pressure, in the sense that
it is due to the bounded motion of virtual electron-positron pairs
. For small fields, of order $10-10^3$G, it is negligible as
compared with the usual Casimir pressure. But for larger fields,
e.g. for $B\sim 10^{5}$ G it becomes larger; one may obtain then
pressures up to $P_{V \perp} \sim10^{-9} dyn/cm^{2}$ (Fig.
\ref{fig:casimir}). For a distance between plates $d=0.1cm$, gives
$P_C \sim 10^{-14}dyn/cm^{2}$ , i.e., five orders of magnitude
smaller than $P_{V \perp}$. This suggests that vacuum interaction
with the magnetic field may produce observable effects for fields,
which can be realized in nature or in terrestrial laboratories.

\section{Vacuum pressures for strong fields}

 \indent

 At this point, thinking in possible astrophysical consequences, it is useful to
analyze the behavior of transverse vacuum pressure for large
magnetic fields, i.e., for fields of the order of the critical one
$B_{c}=m^{2}_{e}/e\sim10^{13}G$ and larger. Fields of these orders
can be generated due to gravitational and rotational effects in
stellar objects \cite{chak}, where the electron-positron gas play
an important role. We use a model of white dwarf, in which the
main contribution to thermodynamic magnitudes comes from the
electron sector, described as a degenerate quantum gas. There is
also a nuclei background, which compensate the electrical charge,
but it behaves like a classical gas, and leads to quantities
negligible small as compared with those of the electron gas.

The statistical contribution of electrons to the transverse
pressure, in the degenerate limit, is
\begin{equation}\label{stpressure}
P_{ST \perp}=\frac{2(eB)^{2}}{\pi^{2}}
\sum_{n=1}^{n_{\mu}}n\ln\frac{\mu+\sqrt{\mu^{2}-m^{2}-2eBn}}{\sqrt{m^{2}+2eBn}},
\end{equation}
where $n$ is the Landau state number, and $\mu$ is the chemical
potential, related to the electron density through the expression
\begin{equation}\label{mu}
N=\frac{eB}{2\pi^{2}}\sum_{n=0}^{n_{\mu}}\alpha_{n}\sqrt{\mu^{2}-m^{2}-2eBn},
\end{equation}
and $n_{\mu}$ is an integer
\begin{equation}
n_{\mu}=I(\frac{\mu^{2}-m^{2}}{2eB}).
\end{equation}

In the extreme case of sufficiently strong magnetic field
$2eB>\mu^{2}-m^{2}$, all the electron system is in the Landau
ground state $n=0$. But it is easy to note that this state does
not contribute to the statistical pressure. In other words, when
the electrons are confined to the Landau ground state, the total
transverse pressure is equal to the vacuum one and is negative. It
causes system instability. It means that there is a limit value of
magnetic field for which the system stability may be preserved,
and it depends on the electron density
\begin{equation}\label{Blim}
  B_{lim}=AN^{2/3},   A=\frac{(2\pi^{4})^{1/3}}{e}.
\end{equation}
For fields $B=B_{lim}$ the instability is produced, and the
transverse pressure becomes negative due to the vacuum term. For
less fields, that is for $B<B_{lim}$, we must consider both
contributions: statistical and vacuum one. For typical densities
of a white dwarf $N\sim10^{30}electrons/cm^{3}$  we find
$B_{lim}\sim10^{13}G$. In this particular case, if $B<B_{lim}$ we
get $P_{\perp}\approx P_{ST \perp}\sim10^{23}-10^{24}dyn/cm^{2}$,
and $P_{ \perp}= P_{V \perp}\sim10^{19}-10^{22}dyn/cm^{2}$ for
$B=B_{lim}$. We conclude that when $B=B_{lim}$ the star collapses.
It agree with the results obtained in \cite{h3}.

 A different situation is produced for larger fields. For
$B\sim10^{15}G$ , pressures of order  $P_{V
\perp}\sim10^{25}-10^{28}dyn/cm^{2}$ are found, and thus,
\textit{it may be of the same order than the statistical term},
although densities $N\sim10^{33}electrons/cm^{3}$ are required to
keep $P_{ST \perp}\neq 0$. The vacuum contribution becomes of the
same importance of the statistical one and must be taken into
account in any analysis of the system stability. In fact for these
field intensities , it may happen that the total transverse
pressure  may not vanish, and even become negative, although the
electron system were not confined to the Landau ground state
$n=0$.

\section{Ultra strong fields $eB>>m^{2}$ }

\indent

In the previous section we have seen that the effects of vacuum
interaction with the magnetic field can not be neglected if we
compare them with the effects produced due to the same interaction
of the real particle system, for $B\sim10^{15}G$. We can expect
then, that the vacuum role becomes more and more relevant for
larger fields. One way of handling this problem is by including
the effect of radiative corrections through an anomalous magnetic
moment, and modifying consequently the energy spectrum, which
would appear as a solution of the Dirac equation for a charged
particle with anomalous magnetic moment \cite{anom moment}
\begin{equation}\label{anom energy}
  \varepsilon_{n,\eta}^{a}=\sqrt{p_{3}^{2}+
  (\sqrt{m^{2}+(2n+\eta+1)eB}+\eta\frac{\alpha}{2\pi}\mu_{B}B)^{2}},
\end{equation}
for $ n=0,1,2,...$, where $\mu_{B}$ is the Bohr magneton, and
$\eta=1,-1$ are the $\sigma_{3}$ eigenvalues corresponding to the
two orientations of the magnetic moment with respect to the field
$B$. The vacuum thermodynamic potential has now the form
\begin{equation}\label{pot2}
 \Omega_{V}=-\frac{eB}{4\pi^{2}}\sum_{\eta=1,-1}\sum_{n=0}^{\infty}
\int_{-\infty}^{\infty}dp_{3}\varepsilon_{n,\eta}^{a}.
\end{equation}
After regularization, it transforms in to
\begin{equation}\label{pot2.1}
  \Omega_{V}=\frac{(e
B)^{2}}{8\pi^{2}}\sum_{k=0}^{\infty}\frac{4^{k}}{(2k)!}
((\frac{\alpha}{4\pi})^{2}\frac{eB}{m^{2}})^{2k}\Lambda_{k},
\end{equation}
where
\begin{equation}\label{pot terms}
 \Lambda_{0}= \int_{0}^{\infty}\frac{dx}{x^{2}}
e^{-(\frac{m^{2}}{eB}+(\frac{\alpha}{4\pi})^{2}\frac{eB}{m^{2}})x}(\coth
x-\frac{1}{x}-\frac{1}{3}x),
\end{equation}
and
\begin{equation}\label{pot terms2}
   \Lambda_{k}= \int_{0}^{\infty}dx
e^{-(\frac{m^{2}}{eB}+(\frac{\alpha}{4\pi})^{2}\frac{eB}{m^{2}}+1)x}
(\frac{1}{\sinh x}-\frac{1}{x})
((\frac{4\pi}{\alpha})^{2}\frac{m^{2}}{eB}\frac{d}{dx}-1)^{k}x^{2(k-1)},
\end{equation}
for $k\neq0$. Due to the smallness of the factor
$\frac{\alpha}{4\pi}\sim10^{-4}$, in the series expansion
(\ref{pot2.1}) we can neglect all terms $\Lambda_{k}$ with
$k\neq0$, except for fields
$B\sim10^{-1}(\frac{4\pi}{\alpha})^{2}B_{c}$, being
$(\frac{4\pi}{\alpha})^{2}B_{c}\sim10^{20}G$, when more terms
should be include.
\begin{figure}[t]
\epsfxsize=25pc 
\epsfbox{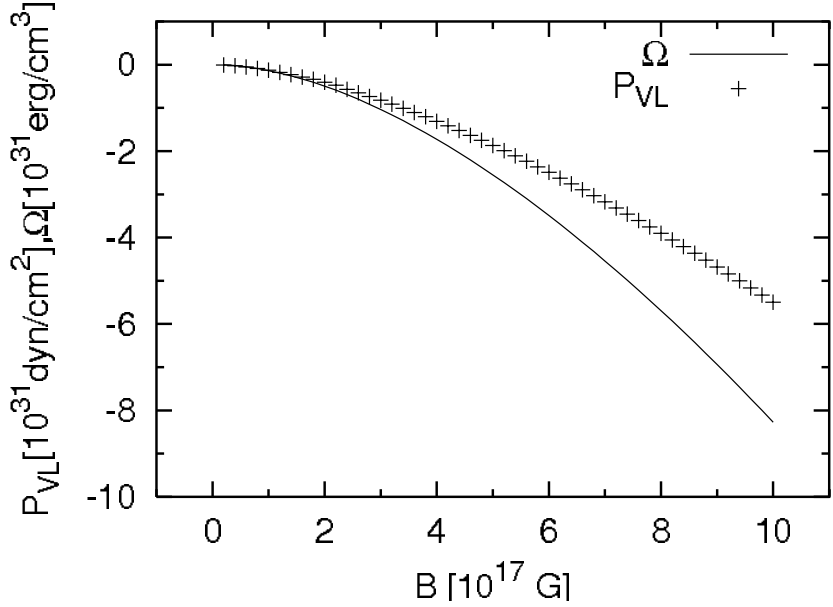} 
\caption{Vacuum transverse pressure and energy dependence on the
field, taking into account the electron anomalous magnetic moment.
\label{fig:pr.anom}}
\end{figure}

The magnetization and transverse pressure become
\begin{equation}\label{an.mag}
  {\cal M}_{V}=-\frac{2\Omega_{V}}{B}-
\frac{e^2B}{8\pi^2}(\frac{m^{2}}{eB}-(\frac{\alpha}{4\pi})^{2}\frac{eB}{m^{2}})
\int_0^{\infty}\frac{dx}{x}
e^{-(\frac{m^{2}}{eB}+(\frac{\alpha}{4\pi})^{2}\frac{eB}{m^{2}})x}
(\coth x-\frac{1}{x}-\frac{1}{3}x),
\end{equation}
\begin{equation}\label{an.pressure}
  P_{V \perp}=\Omega_{V}+
\frac{(eB)^2}{8\pi^2}(\frac{m^{2}}{eB}-(\frac{\alpha}{4\pi})^{2}\frac{eB}{m^{2}})
\int_0^{\infty}\frac{dx}{x}
e^{-(\frac{m^{2}}{eB}+(\frac{\alpha}{4\pi})^{2}\frac{eB}{m^{2}})x}
(\coth x-\frac{1}{x}-\frac{1}{3}x).
\end{equation}
It must be noted that for
$B>\frac{4\pi}{\alpha}\frac{m^{2}}{e}\sim10^{16}G$ although
$p_{\perp}$ remains  negative, it grows more slowly than for
fields $B<\frac{4\pi}{\alpha}\frac{m^{2}}{e}$, and its absolute
value is smaller than $p_{3}$ (Fig. \ref{fig:pr.anom}).

\section{A quark gas}

\indent

Fields of very high orders $B\sim10^{17}-10^{20}G$ have been
suggested to be found in the cores of extremely magnetized neutron
stars. But an electron system is hardly in equilibrium for this
range of fields, and for these stellar objects the neutron and
proton gases are the most important. The last suggests  that the
interaction of quark-antiquark virtual pairs with the magnetic
field should be studied. As the interaction energy with the
magnetic field may become comparable with the color field
interaction among quarks,
 we
may consider the quark interaction with the magnetic field in a
similar way as the electron gas. We may write the electromagnetic
vacuum energy of the quark-antiquark field in presence of the
magnetic field. The main contribution comes from the u-type quark,
which has a smaller mass than other quarks. For u-type quark we
can write

\begin{equation}\label{upot}
\Omega_{V}^{u}=-\frac{(q_{u}B)^{2}}{8\pi^{2}} \int_{0}^{\infty}
e^{-(\frac{m_{u}^{2}}{q_{u}B}+(\frac{\alpha_{u}}{4\pi})^{2}\frac{q_{u}B}{m_{u}^{2}})x}(\coth
x-\frac{1}{x}-\frac{1}{3}x)\frac{dx}{x^{2}},
\end{equation}
where $q_{u}=\frac{2}{3}e$ and $m_{u}$ are the u-quark charge and
mass, respectively, and $\alpha_{u}=q_{u}^{2}/\hbar c$. The
expression (\ref{upot}) for the vacuum quark-antiquark
electromagnetic energy is valid for fields up to $B\sim10^{19}G$.
The magnetization has the form
\begin{eqnarray}
 \nonumber {\cal M}^{u}_{V}=-\frac{2\Omega_{V}^{u}}{B}-
\frac{q_{u}^2B}{8\pi^2}(\frac{m_{u}^{2}}{q_{u}B}
-(\frac{\alpha_{u}}{4\pi})^{2}\frac{q_{u}B}{m_{u}^{2}})
\int_0^{\infty}
e^{-(\frac{m_{u}^{2}}{q_{u}B}+(\frac{\alpha_{u}}{4\pi})^{2}
\frac{q_{u}B}{m_{u}^{2}})x}\times \\ \times(\coth
x-\frac{1}{x}-\frac{1}{3}x )\frac{dx}{x},
\end{eqnarray}
and the transverse pressure is
\begin{eqnarray}\label{uan.pressure}
  \nonumber p_{V \perp}^{u}=\Omega_{V}^{u}+
\frac{(q_{u}B)^2}{8\pi^2}(\frac{m_{u}^{2}}{q_{u}B}
-(\frac{\alpha_{u}}{4\pi})^{2}\frac{q_{u}B}{m_{u}^{2}})
\int_0^{\infty}
e^{-(\frac{m^{2}}{q_{u}B}+(\frac{\alpha_{u}}{4\pi})^{2}\frac{q_{u}B}{m_{u}^{2}})x}
\times \\ \times(\coth x-\frac{1}{x}-\frac{1}{3}x )\frac{dx}{x}.
\end{eqnarray}

\begin{figure}[t]
\epsfxsize=25pc 
\epsfbox{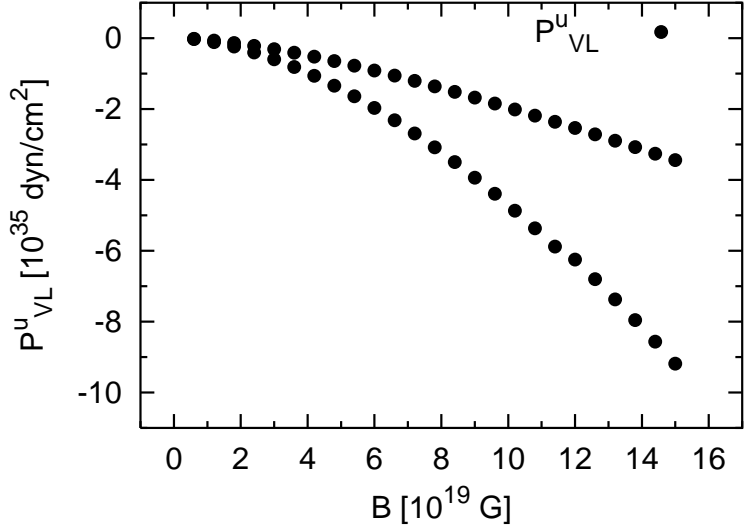} 
\caption{Transverse pressure exerted by the u-type quark-antiquark
virtual pairs. \label{fig:upressure}}
\end{figure}

The transverse vacuum pressure gives figures of the order of $p_{V
\perp }^{u}\sim 10^{29}-10^{31}dyn/cm^{2}$ for fields $B\sim
10^{17}G$, and $p_{V \perp}^{u}\sim 10^{33}-10^{35}dyn/cm^{2}$ for
fields $B\sim 10^{19}G$, as is shown in Fig. \ref{fig:upressure}.
\section{Conclusions}

\indent

A magnetic field modify the  electron-positron zero-point energy
of vacuum, leading to new physical effects. The vacuum shows a
nonlinear response, as a ferromagnetic medium. A negative
pressure, exerted in the direction perpendicular to the field,
appears as a Casimir- like force, which must produce observable
effects for fields $B\sim 10^{5}G$.  The electron-positron
zero-point energy leads to a transverse pressure of similar order
of the statistical one for fields $B\sim10^{15}G$, and densities
$N\sim10^{33}electrons/cm^{3}$ or larger would be required in
order to avoid the collapse.

Vacuum shrinks perpendicular to the magnetic field. This leads us
to conclude that nuclear and/or quark matter is unstable for
fields $B\sim m_{n}^{2}/e$, $m_{n}$ being the nucleon mass. As the
contribution of the transverse quark pressure is smaller than its
longitudinal term, the vacuum negative transverse pressure makes
the system to implode.

\section*{Appendix}

\indent

We start from the expression for $\Omega_{V}$ (\ref{pot2}). Using
the integral representation
\begin{equation}\label{rep}
  a^{1/2}=\frac{1}{2}\pi^{-1/2}\int_{0}^{\infty}dtt^{-3/2}(1-e^{-at}),
\end{equation}
 and taking $a^{1/2}=\varepsilon_{n,\eta}^{a}$, we can perform the Gaussian
 integral on $p_{3}$. We obtain
\begin{equation}
 \Omega_{V}(\epsilon)=\frac{eB}{8\pi^{2}}\sum_{\eta=1,-1}\sum_{n=0}^{\infty}
 \int_{\epsilon}^{\infty}\frac{dt}{t^{2}}
 e^{-(\sqrt{m^{2}+(2n+\eta+1)eB}+\eta\frac{\alpha}{4\pi}\frac{eB}{m})^{2}t},
\end{equation}
where we have introduced a quantity $\epsilon$ in order to
regularize the divergent term dependent on $a$ in (\ref{rep}).
Performing the sum over $\eta$, we can write
\begin{eqnarray}
 \nonumber \Omega_{V}(\epsilon)=\frac{eB}{8\pi^{2}}
  \int_{\epsilon}^{\infty}[e^{-(m-\frac{\alpha}{4\pi}\frac{eB}{m})^{2}t}
 +2\sum_{n=1}^{\infty}
 e^{-(m^{2}+2neB+(\frac{\alpha}{4\pi}\frac{eB}{m})^{2})t}\times\\
 \times \cosh \frac{\alpha}{2\pi}\frac{eB}{m}t\sqrt{m^{2}+2neB}]\frac{dt}{t^{2}}.
\end{eqnarray}
Substituting the series expansion of $\cosh
\frac{\alpha}{2\pi}\frac{eB}{m}t\sqrt{m^{2}+2neB}$, we get
\begin{equation}
  \Omega_{V}(\epsilon)=\frac{eB}{8\pi^{2}}
 \int_{\epsilon}^{\infty}\frac{dt}{t^{2}}
 (e^{-(b_{0}-g)^{2}t}+
 2\sum_{n=1}^{\infty}\sum_{k=0}^{\infty}\frac{(2gb_{n}t)^{2k}}{2k!}
 e^{-(b_{n}^{2}+g^{2})t}),
\end{equation}
where $b_{n}=\sqrt{m^{2}+2neB}$ and
$g=\frac{\alpha}{4\pi}\frac{eB}{m}$. Using the fact that
\begin{equation}
 b_{n}^{2}\int_{\epsilon}^{\infty}dte^{-(b_{n}^{2}+g^{2})t}t^{2(k-1)}=
\int_{\epsilon}^{\infty}dte^{-(b_{n}^{2}+g^{2})t}[\frac{d}{dt}-g^{2}]t^{2(k-1)},
\end{equation}
and performing the sum over
 $n$, we find that
\begin{eqnarray}\label{pot.fin}
 \nonumber \Omega_{V}(\epsilon)=\frac{eB}{8\pi^{2}}
 \int_{\epsilon}^{\infty}\frac{dt}{t^{2}}
 (e^{-(b_{0}-g)^{2}t}+
 2\frac{eB}{8\pi^{2}}\sum_{k=0}^{\infty}\frac{(2g)^{2k}}{2k!}\int_{\epsilon}^{\infty}dt
 e^{-(b_{0}^{2}+g^{2})t}\times \\
 \times e^{-eBt}\frac{1}{\sinh
 eBt}[\frac{d}{dt}-g^{2}]x^{2(k-1)}).
\end{eqnarray}
Due to the smallness of $\frac{\alpha}{2\pi}\sim10^{-4}$, we can
approximate in (\ref{pot.fin}) $1\pm\frac{\alpha}{2\pi}\approx1$
and write
\begin{equation}\label{aprox}
  e^{-2b_{0}gt}+e^{-eBt}\frac{1}{\sinh eBt}\approx
  \coth eBt.
\end{equation}
Finally, subtracting to $\coth eBt$ and $\frac{1}{\sinh
 eBt}$ the first terms in their series expansion, we can take
 $\epsilon\rightarrow0$, and obtain the expression (\ref{pot2.1}),
 where we put $eBt=x$.


\small
\begin{thebibliography}{000}
\bibitem{Casimir} H.B.G. Casimir, Proc. K. Ned. Akad. Wet. \textbf{51}, 793
(248)
\bibitem{Milonni}P.W. Milonni, The Quantum Vacuum, Acad. Press,
New York (1994)
\bibitem{Euler} W. Heisenberg and
H. Euler, Z. Phys. 98, 714 (1936)

\bibitem{h3}  M. Chaichian, S. Masood, C. Montonen, A. P\'{e}rez
Mart\'{i}nez, H. P\'{e}rez Rojas, Phys. Rev. Lett. {\bf 84\ }5261{\bf \ }%
(2000).

\bibitem{h plasma}  H. P\'{e}rez Rojas,{\it \ Acta Phys. Pol. }{\bf B 17}
861 (1986).
\bibitem{Aurora} A. P\'{e}rez
Mart\'{i}nez, H. P\'{e}rez Rojas,H. J. Mosquera Cuesta, Eur. Phys.
Jour. C, 29, 111-123 (2003)
\bibitem{Jackson}  J. D. Jackson, Classical Electrodynamics, J. Wiley and Sons,
 New Jork, (1963)
\bibitem{Boyer}  T. H. Boyer,{\it \ Phys. Rev.}{\bf 174} 1764-1777 (1968).
\bibitem{chak}  S. Chakrabarty, D. Bandyopadhyay, and S. Pal,{\it \ Phys. Rev. Lett.}
{\bf 78} 2998 (1997).
\bibitem{anom moment}  H. Y. Chiu, V. Canuto and L. Fassio-Canuto,{\it \ Phys.
Rev.}{\bf 176}
 1438-1443 (1968).

\end{thebibliography}
\end{document}